# Topological Signatures in the Electronic Structure of Graphene Spirals

Stas.M. Avdoshenko,[1,*] Pekka Koskinen,[2] Haldun Sevinçli,[3] Alexey A. Popov,[4] and Claudia G. Rocha[2]

[1]*School of Materials Engineering, Purdue University, West Lafayette, Indiana, USA*

[2]*Nanoscience Center, Department of Physics, University of Jyväskylä, 40014 Jyväskylä, Finland*

[3]*DTU Nanotech, Department of Micro- and Nanotechnology, Technical University of Denmark, DK-2800, Denmark*

[4]*The Leibniz Institute for Solid State and Materials Research in Dresden, 01069 Dresden*

(Dated: November 9, 2012)

## Abstract

Topology is familiar mostly from mathematics, but also natural sciences have found its concepts useful. Those concepts have been used to explain several natural phenomena in biology and physics, and they are particularly relevant for the electronic structure description of topological insulators and graphene systems. Here, we introduce topologically distinct graphene forms - graphene spirals - and employ density-functional theory to investigate their geometric and electronic properties. We found that the spiral topology gives rise to an intrinsic Rashba spin-orbit splitting. Through a Hamiltonian constrained by space curvature, graphene spirals have topologically protected states due to time-reversal symmetry. In addition, we argue that the synthesis of such graphene spirals is feasible and can be achieved through advanced bottom-up experimental routes that we indicate in this work.

PACS numbers:

---

[*]Electronic address: `savdoshe@purdue.edu`

## A. Introduction

In mathematics, topology analyzes how the properties of objects preserve under continuous deformations. But the interest to topological analysis is not restricted to mathematics alone; it spans also through biology, chemistry and materials science. In protein systems topology determines when protein folding sustains the rest of their cellular life.[1] In condensed matter physics, topology dominates several quantum phenomena, such as quantum-Hall[2], spin-Hall[3], and Aharonov-Bohm[4] effects, as well as the physics of topological insulators[5].

In topological insulators, the surface electronic states are governed by topological features, making their quantum information robust against impurity scattering. Such robustness, by being protected by time-reversal-invariant Hamiltonian, could pave a reliable avenue toward fault-tolerant quantum-computing technology.[6] Experiments via angle-resolved photoemission spectroscopy performed in $Bi_2Se_3$ compounds[7] and $Bi_{1-x}Sb_x$ alloys[8] have shown signatures specific to topological insulators, such as large bulk energy gap and a single-surface Dirac cone associated to its topologically protected state. Dirac cones make the physics of graphene and topological insulators similar, even though graphene has two Dirac valleys with spin degeneracy while topological insulators have only one Dirac valley without spin degeneracy.[9] In addition, graphene can exhibit topologically protected quantum-Hall states with applied perpendicular and periodic magnetic fields.[10]

In the absence of *structure inversion symmetry*, surface states may split because of Rashba spin-orbit interaction.[11] This splitting has been verified in thin films[12] and semiconductor heterostructures having an inversion asymmetry of the confining potentials.[13–15] The splitting arises from the Rashba interaction Hamiltonian, $H_R = \alpha(\vec{E} \times \vec{p}) \cdot \vec{s}$, where $\alpha$ is the Rashba coefficient, $\vec{s}$ is the spin of an electron moving with momentum $\vec{p} = \hbar\vec{k}$ in an electric field $\vec{E}$.[16] In a two-dimensional non-interacting electron gas, therefore, Rashba spin-orbit interaction splits the parabolic energy bands in two, $\epsilon_\pm(k) = \hbar^2 k^2/2m^* \pm \alpha k$, where $m^*$ is the effective mass.[17] Even though spin-orbit interactions are often intrinsic, such as the spin-orbit-induced $\sim$ 0.1 meV energy gap in graphene[18], Rashba splitting is interesting for applications because the control over an external electric field makes it extrinsic and tunable.[19]

Extrinsic spin-orbit manipulation in graphene has been probed by external elec-tric fields[20], by doping[21], by mechanical folding[22] and by depositing graphene on substrates[23]. Especially substrate interfaces, by always involving inversion asymmetries, strengthen Rashba interaction.[24] In graphene-based structures, however, the origin of the Rashba interaction is unlike in any other material: the interaction, although similar to the usual spin-orbit effects, arises not from the real electron spin, but from the spin related to the two non-equivalent atomic sites in the unit cell, the pseudospin. In graphene the Rashba splitting occurs around $k$-points displaying time-reversal symmetry[25], meaning that the splitting is seen around Brillouin zone centers and zone boundaries.[26] In graphene the zone boundary points $K$ and $K'$ are non-equivalent, but they have the same energy and they are connected by time-reversal symmetry.[27, 28] This symmetry makes the connection to topological insulators. The symmetry protects a pair of gapless helical edge modes in topological insulators belonging to $Z_2$ class.[29] In a prototype of such topological insulator, an intrinsic spin-orbit coupling induces a topological mass term in the electronic structure of an atomic hexagonal frame.[30, 31]. Also, it is important to mention that most often $Z_2$ class materials are discussed, at least experimentally, including heavy elements with strong spin-orbital coupling like $Bi_2(Se,Te)_3$ compounds[7].

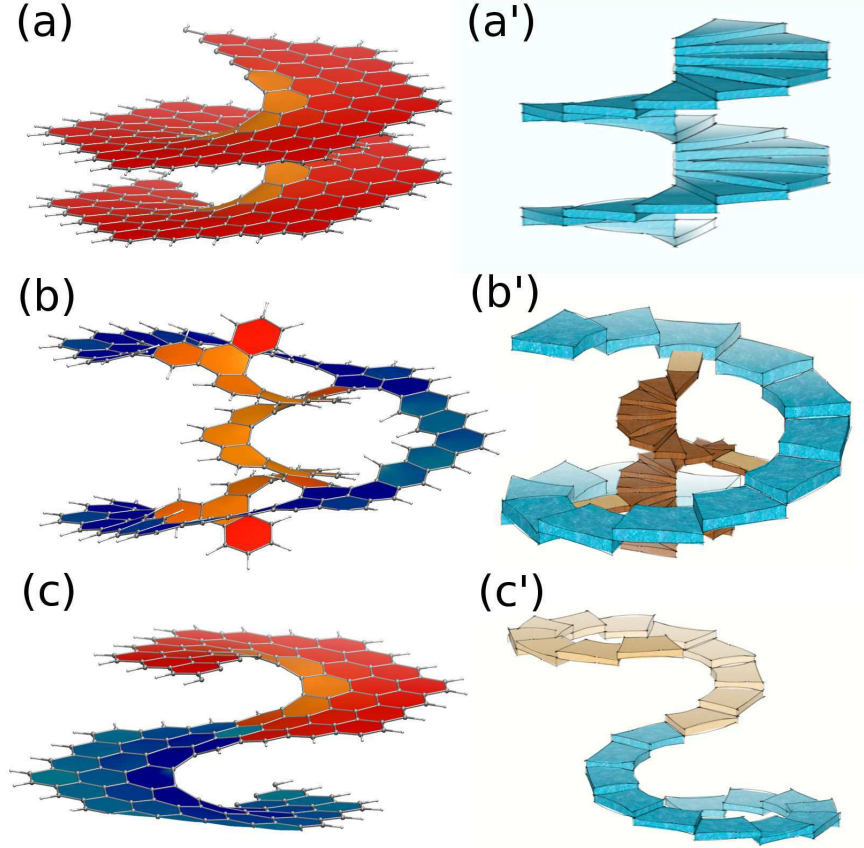

FIG. 1: Examples of graphene spirals. (a) Right-handed spiral, (b) interconnected right- and left-handed double spiral, and (c) loxodrome-like spiral. (a'-c') Mapping of the (a-c) structures into helical stairs to highlight the topology of their curved space.

Till now spin-orbit effects with highly protected topological edge-states in graphene have been obtained only by external perturbations such as external fields, heavy mechanical distortions, or chemical doping. In this letter, therefore, we investigate the above-discussed topological signatures in distinct graphene systems: the graphene spirals (see Fig.1). As a central result, we find out that the spiral topology creates an intrinsic Rashba splitting. While in canonical illustrations of the effect in which an external magnetic field is applied to induce electron precession, in graphene spirals the track in which electrons move already displays an helical topology; electrons are constrained to move along an helical path in k-space and, due to unbreakable structure inversion symmetry, the Rashba-like band structure topology becomes an intrinsic effect for this material class. Our results demonstrate that graphene spirals naturally possess robust topological states as those observed in topological insulators.

### B. Structure and Topology of Graphene Spirals

Graphene spirals are distinct from the helical graphene motifs reported earlier, such as graphene stripes and ribbons bent to spiral-like shapes.[32–35] In those earlier motifs, the starting point has been a graphene ribbon itself, with regular edges and curvature- or strain-modified $\pi$-electron system.[36, 37] In such systems there has always been one-to-one mapping between the helical structure and flat graphene. In other words, it is always possible to build such systems by cutting them out of an infinite and flat graphene sheet. On the contrary, in graphene spirals such a mapping does not exist (see Fig.1). Spirals are one-dimensional systems, while they still have a graphite layered structure containing perfect hexagons. Edge profiles alternate between armchair and zigzag shapes. Since spirals' local structures resemble graphene, they facilitate chiral topology without overly perturbing the $\pi$-electron system. The largest perturbations take place at the inner edge of spirals where the strain is largest.

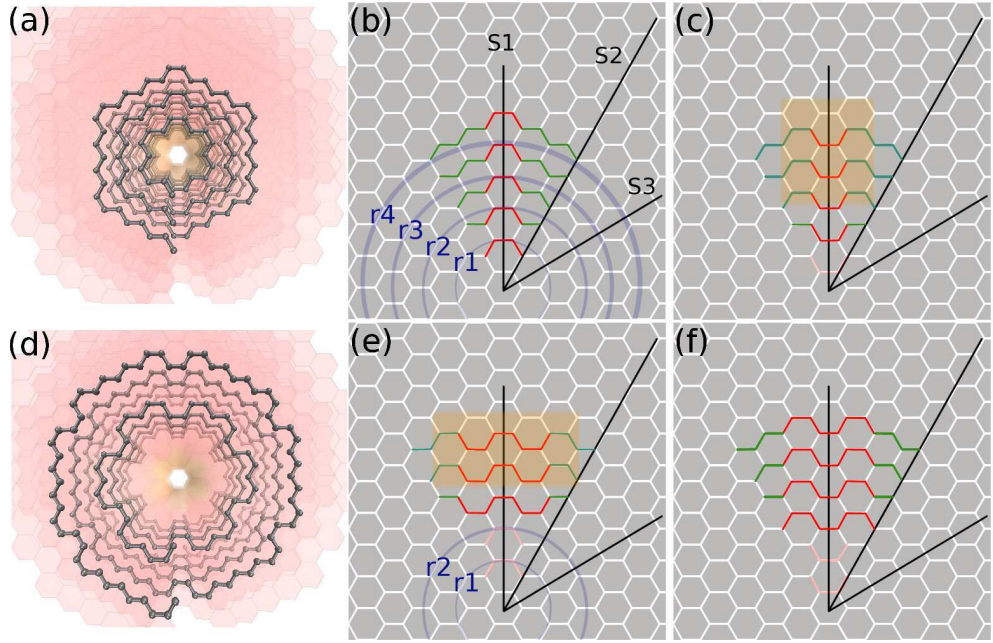

FIG. 2: Notation of graphene spirals edges: Right, Middle and left panels depict three-dimensional (3D) top view, conventional, and mirrored spirals with flipped ac segments, respectively. Note that on the right panels only carbon-carbon bonds along the inner and outer edges of the spirals are highlighted. Armchair (ac) segments are marked in red while zigzag (zz) segments are drawn in green colour. S1, S2 and S3 black lines are the hexagonal symmetry axes ($\pi/6$ wedge angles) and pass through the origin point located on the center of the spiral. The blue circles outline possible extensions that the spirals can hold along their inner and outer edges. (a) 3D view of ac[1]zz[0-4], with its planar projection shown in (b). (c) Planar projection of ac*[1]zz[0-4] spiral where ac segments are flipped. (d) 3D view of ac*[2]zz[1-3], with its planar projection shown in (f). (e) Planar projection of ac[2]zz[1-3] spiral.

First, we illustrate a notation scheme that we elaborated to identify the graphene spiral models (see Fig. 2). This notation is based on the number of armchair (ac) and zigzag (zz) segments used to complete a full coil of the spiral along the inner and outer edges. We established that the spiral axis goes through the center of a hexagon where $\pi/6$-symmetry lines merge. Distinct spiral classes can be formed by combining ac and zz segments. The logic of our notation scheme can be better illustrated by projecting the spiral systems onto the plane. According to our construction rules, the spirals can only be formed by hexagons, their edges can simply contain ac- and zz- fragments and there must be no untangled bonds.

How wide is the spiral can be determined by the outlined circles with radii $r_1$, $r_2$, ..., $r_j$ being $r_j$=1.23+2.46$j$ Å. These circles always enclose the first series of ac segments crossing the symmetry line S1. The remaining of the spiral is completed by as much zz segments required to maintain the perfect hexagonal frame. In this way, we define the notation ac[$m$]zz[$n_{in}$-$n_{out}$], where $m$ is the number of ac units and $n_{in/out}$ is the number of zz units used to complete, respectively, the spiral inner and outer edges. Spirals with ac segments which are flipped towards the origin have an additional "*" in their notation, e.g. ac*[2]zz[1-3] [see Fig.2 (c) and (d)]. These conventions impose the spirals to be invariant under an axial translation of $b$ (interlayer separation), under an axial translation of $b/6$ combined with a rotation of $\pi/3$. Furthermore, it is worth mentioning that almost any parametric curve can be used to represent edges formed under such hexagonal basis, however analysis involving more complex edge topologies are far beyond the scope of this work. The geometry and electronic structure of the built graphene spirals were investigated within density functional theory (DFT) implemented within SIESTA package[38]. To confirm our results, SIESTA calculations were compared to other methods such as single $\pi$-band tight-binding (see Supplementary material), density-functional tight-binding implemented within DFTB+ code[39], and VASP density-functional package[40]. Detailed description of the used parameters and calculation conditions are presented in the end of the manuscript. We obtained the band structures for distinct graphene spiral models which had their atomic configuration fully optimized. The results are discussed in the following.

### C. Rashba Effect in the Electronic Structure of Graphene Spirals

*Ab initio* band structures obtained for graphene spirals of different widths are shown in Fig. 3. One can see that the two narrower spirals are semiconductors with an energy gap of approximately 2.3 and 1.0 eV for ac[1]zz[0] and ac[1]zz[0-1], respectively. Differently, ac[1]$zz$[0-2] and ac[1]$zz$[0-3] reveal semimetallic states. However, more intriguing than the change of electronic character in response to geometrical variations is the peculiar band splitting at the Brillouin zone center. These band splittings represent the Rashba effect, the topological signatures in the electronic structure of graphene spirals. This is our central result. The splitting is robust and present in all spirals. The origin for the Rashba effect is related to how the Dirac particles in graphene couple with the spiral curved space. Graphite crystals, for instance, already manifest a natural splitting of the π-bands close to the Fermi level due to the two non-equivalent carbon atoms in the unit cell. Since the number of atoms in the spiral unit cell - assumed to be an unique coil - is considerably higher, one expects a superior number of splittings in these systems. In addition, the interlayer interaction in spirals also plays an important role. The usual graphite layer stacking provides an uniform potential profile over the whole sample which is not the case for the spirals. They possess chiral symmetry which leads to electron-hole symmetry breaking. Under chiral symmetry operation $\hat{S}_\theta^\tau$, the wavefunction composed of a set of molecular orbitals within the unit cell get translated (by $\tau$) and rotated (by $\theta$) simultaneously[41]. The operator $\hat{S}_\theta^\tau$ replaces the translation operator $\hat{T}$ in Bloch's theorem[42]. It is therefore important to distinguish reciprocal k-vectors of the linear ($k$) and curved or chiral ($k'$) systems. These vectors are related by $k' = (b/\tau)k$, as can be demonstrated by applying $\hat{S}$ consecutively until a full turn is completed. The enlarged Brillouin zone in chiral systems is often referred to the Jones zone (JZ)[43]. It can be shown that in a constantly curved space systems, a *bisector* reduction of whole the Brillouin space must have two components as for any system described by a Hamiltonian under chiral symmetry.

Let us now illustrate in detail the origin of the Rashba effect in our graphene spirals by considering a linear chain model as starting point (see also Supplementary material). The chain periodicity is given by the unit cell length $a$, and chain's one-dimensional potential obeys $V(z) = V(z+a)$. When the unit cell gets more atoms, the reciprocal space shrinks, the bands fold and their number increases. For unit cells with an even number of atoms, two types of bands exist: bands crossing the Fermi level (i) at the $\Gamma$ point and (ii) at the edges of the Brillouin zone. At the $\Gamma$ point, the energy states are quantized as for a finite ring with $M$ atoms, $\epsilon_n(k) = -2t\cos(2\pi n/M)$, where $n = 1, \ldots, M$ and $t$ is the hopping parameter. Only when $2n/M = \pm 1/2, \pm 3/2, \pm 5/2, ...$, a band will cross the Fermi level at $\Gamma$ point. For unit cells with an odd number of atoms, the band crossing is shifted from $\Gamma$ point and an energy gap opens. A reminiscence of this cosine trend exists also in polyacetylene-like spiral (not shown). For spirals with large circumference arcs, the low-energy band slopes at the edges of the JZ are markedly higher, meaning curvature-dependent Fermi-velocities. This dependence is absent in single-π band tight-binding model since it cannot account for curvature effects. Furthermore, in these spirals the chiral operation changes the orbital orientation. The angular parts of the p-orbitals can be written as $p_x(l) = p_x \cos(l\theta)$, $p_y(l) = p_y \sin(l\theta)$, and $p_z(l) = p_z$, where $l$ is an integer number for consecutive units cells of two non-equivalent carbon atoms. This orbital orientation changes the phase of the wavefunction, a $180^o$ turn flips its sign; this property modifies the profiles of bonding and antibonding bands markedly, due to additional cross-coupling terms[43, 44]. In the simpler spiral example shown on Fig. 3(a), its valence band develops minima at X-points and maxima around the $\Gamma$-point. As the spiral becomes wider, the electronic structure becomes richer but still such peculiar band offset is maintained. This energy uplift and concomitant maximum development around the $\Gamma$ point arises because the states become more antibonding due to chiral symmetry.

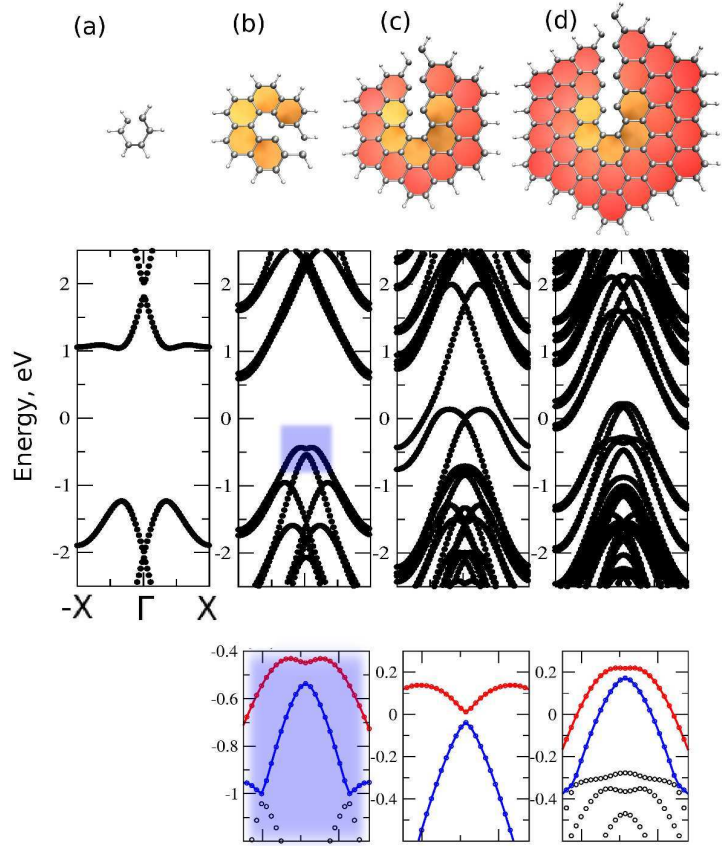

FIG. 3: Band structures of graphene spirals. (a) ac[1]zz[0], (b) ac[1]zz[0-1], (c) ac[1]zz[0-2], and (d) ac[1]zz[0-3]. Fermi energy is set at 0 eV. Vertical axes span the $k$-vector intervals $[-\pi/b,\pi/b]$. Narrow spirals, semiconductors with $2.3 - 1.0$ eV energy gaps; wide spirals, semimetals. Bottom panels depict a zoom over low energy bands around the $\Gamma$. The band closest to the Fermi energy is highlighted in red while the second closest is colored in blue.

Zooming over the low energy bands around $\Gamma$ point [bottom panels of Fig. 3], one can notice that they display intriguing anticrossings close to the Fermi energy as the spirals get wider. This occurs due to certain selection rules (similar to those manifested in ordinary atomic rings) that the helical states must obey combined with the chiral symmetry of the curved space. These anticrossings between such chiral branches can be finely tuned by means of external fields. This is shown on Fig. 4 which depicts the band structure for the spiral ac[1]zz[0-2] while a high intense external electric field is applied perpendicularly its axial direction. At such intense electric fields, one would expect more impacting changes on the electronic structure of the spirals. Nonetheless, such robustness is simply a remarkable proof that these helical modes are protected by time-reversal symmetry. We can allude the nature of these states by assuming that the spirals are composed of $q$ concentric rings embedded in the same unit cell. if the spiral respects hexagonal symmetry, i.e. each ring contains a number of atoms such as $M = 6, 18, 30, 42, ...$, according to a tight-binding description, whenever $q$ is even, an energy gap opens in the electronic structure (see Fig. 2 in supplementary information). Otherwise, metallic edge-states touching the Brillouin zone boundaries will form. The same opening-closing rules for the energy gap fail for the first principle results since they take into account curvature effects. However the low energy states still retain its edge-nature as can be seen from the isosurface plots of the local density of states at the Fermi energy for the spiral ac[1]zz[0-2] (see Fig. 3 in supplementary material). Ultimately, by comparing the results derived from *ab initio* and tight-binding methods, we successfully confirmed that the electronic structure of graphene spirals couples to the helical space backbone.

Previous works have reported that ripples in graphene can be modeled by coupling the Dirac equation to a curved metric space defined phenomenologically from corrugations observed in experiments[45, 46]. Such covariant formalism gives rise to an effective Hamiltonian where it is possible to identify that the electrons on the deformed space exhibit a new Fermi velocity. The latter is smaller than the flat graphene case. Their findings were confirmed by Raman spectroscopy measurements performed in folded graphene samples[47]. In our spiral examples, electrons propagating along the helical track with stronger bending would then manifest smaller effective Fermi velocities in comparison to those with smoother curvature. This difference can be clearly seen on the band structures of 3 where the low energy bands associated to each spiral have distinct slopes. Such velocity reduction cannot be captured by the single $\pi$-band tight binding approximation since curvature is invisible within this description. Another prominent feature that can be distinguished from the effective covariant

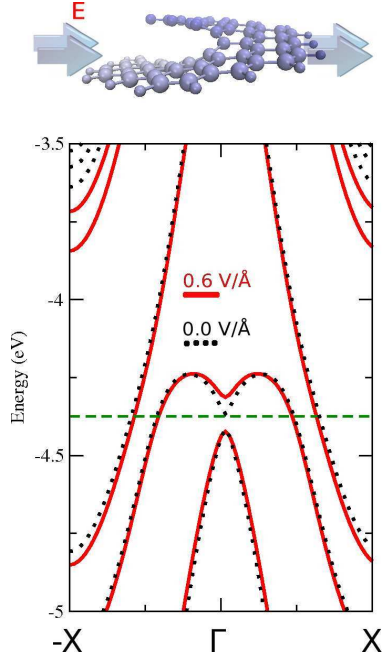

FIG. 4: (red lines) Band structure for the graphene spiral ac[1]zz[0-2] (top panel) under the influence of an external electric field of intensity 0.6 V/Å and applied perpendicularly to its axial direction. Horizontal line marks the Fermi energy (not shifted to zero). Dotted black lines are the energy dispersion for the same spiral for null electric field.

Hamiltonian is the appearance of an effective magnetic field pointing perpendicularly to the deformed graphene sheet[46, 48]. In other words, mechanical strain loaded on the distorted section of the graphene mimics the same effects caused by an external magnetic field (see supplementary information where we demonstrate how the Rashba-like Hamiltonian can be derived from an illustrative toy-model example). The field intensity is strictly related to how sharp the curvature is. This effective field appears naturally from covariant formalism where Dirac equation is solved on curved space. Once the metric of the deformed space is known, this information is plugged on the Dirac equation and techniques resembling to perturbation theory can be used if one assumes that the curvature is rather smooth. The solution for this problem can be often recognized as the standard graphene Dirac model in the presence of an effective potential generated by its own curvature. Such method is very efficient in dealing with smooth ripples or light corrugations on the graphene sheet. Extending such interpretation to our systems, one can already expect that the highly intensified curvature of the graphene spirals will affect enormously their electronic response. Since the curvature of the graphene spirals is so remarkable, the use of covariant methods following expansion procedures is unreliable. In this sense, we must indeed rely on robust *ab initio* methods where all the structures are fully optimized and the effects of the curvature are naturally incorporated in the Hamiltonian.

But even if unreliable, the formalism can help understand the origin of the Rashba splitting. In Aharonov-Bohm (AB) devices real magnetic fields induce electron wavefunctions an additional phase due to the breaking of time-reversal. The phase is proportional to magnetic flux penetrating the device geometry and depends on whether electron moves clockwise or anticlockwise[49]. For the sake of simplicity, let us consider a spiral chain with 18 atoms in its unit cell ($C_{18}$) and investigate its density of states while varying the unit cell length or the curvature of its helix [see Fig.5(b)]. This result was obtained within density functional tight-binding (DFTB) formalism in which the atomic structure of the system-chain could be fully optimized for each unit cell length. Electrons moving clockwise ($k+$) and anticlockwise ($k-$) acquire different phase factors in their wavefunctions depending on the local curvature of the spiral. Thus, an oscillatory pattern akin to AB interference emerges in the density of states [see Fig.5 (a)]. Only here the AB oscillations arise not from a real magnetic field but from a pseudo-magnetic field defined by the system topology; the curvature can be seen as a parameter replacing a real external magnetic field to produce the same effect. As a result, electrons having different pseudospin and being coupled to the pseudo-magnetic field can move in opposite directions and still preserve time-reversal symmetry while exhibiting non-zero AB phase.

### D.  Conclusions

Because of electronic properties similar to topological insulators, graphene spirals could naturally be used in quantum computing. Furthermore, the spiral geometry itself suggests usage as electronics components, as nano-solenoids to produce local magnetic fields. We can estimate the magnetic field intensity, $B$, created by electrons traveling around the spiral of radius $R$. By setting the Lorentz force and the centripetal force equal, we obtain $B = mv_F/eR$, where $m$ is the electron mass, $e$ the electron charge, and $v_F$ the Fermi velocity. Assuming $R \approx 1$ nm and $v_F \approx 10^6$ m/s (Fermi-velocity of flat graphene) we obtain an estimate for graphene spiral -generated magnetic field as $B = 10^3$ T. The estimate depends on the Fermi-velocity, which our results shows to depend on the precise atomic structure of the spiral. The spiral structure can hence be used to sensitively customize the solenoid properties.

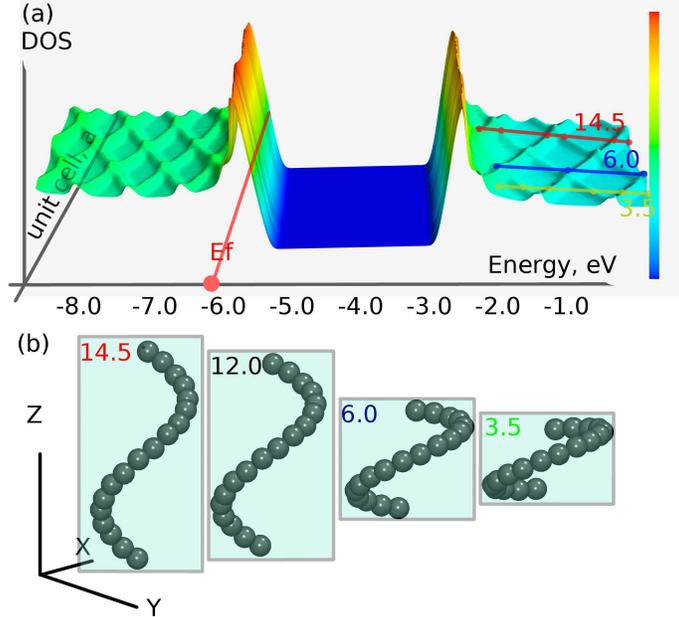

FIG. 5: (a) Aharonov-Bohm-like oscillations in elongated $C_{18}$ atomic chain being gradually deformed into a spiral shape. Normalized density of states *vs.* energy and elongation of the unit cell (in Angstroms). Line at $E_f$, Fermi-level. Horizontal lines highlight the results for some chain geometries that are displayed on panel (b).

Finally, we discuss the feasibility to realize spirals experimentally. We note that natural growth often favors chiral molecules, first and most notable example being the DNA. Second example is carbon nanotubes grown into helical conformations by using controlled creation of pentagon and heptagon defects during the growth process[50, 51]. To synthesize graphene spirals, however, the most relevant experiment was the recent controlled synthesis of narrow graphene nanoribbons.[52] Ribbons were synthesized for controlled ribbon widths, controlled edge shapes and —in particular— controlled topologies. The control was achieved by organic chemical reactions on gold surfaces, where the topology of the nanoribbon was determined by the topology of the precursor monomers. Here, we propose that graphene spirals could be synthesized by the same controlled bottom-up approach. One would only need to choose appropriate precursor monomers, preferably from an organic polymer family with a helical motif. There are also other synthesizing alternatives. Namely, the graphene spiral ac[1]zz[0-1] is already a familiar molecule, the helicene.[53] Fairly long helicenes have been synthesized, and perhaps related techniques could be extended to synthesize also wider graphene spirals.[32, 54] Yet another alternative to spiral synthesis is to use the viewpoint of array of screw dislocations.[51, 55]

To conclude, the electronic structure of graphene spirals show Rashba splitting as a distinct topological signature. The splitting can be understood as a consequence of the intrinsic curvature present in graphene spirals, as a consequence of the coupling between pseudo-spin and the curved helical geometry. The splitting mechanism is similar to the mechanism of *band inversion* in topological insulators.[56, 57] The split, low-energy states around Γ-point are localized at the edges and are protected by the spiral topology, being thus robust against impurities or lattice distortions. These unique electronic properties require neither an external magnetic field nor spin-orbit interaction, which is unlike any typical quantum Hall system. Therefore, graphene spirals ought to deserve a prominent role as a fundamental graphene topology, comparable to the topologies of carbon nanotubes and graphene nanoribbons.

### E. Theoretical Methods

Presented band structure calculations were performed using the SIESTA[38] package within generalized gradient approximation for the exchange and correlation energies[58]. Norm-conserving pseudopotentials[59] with relativistic corrections and a split-valence double-$\zeta$ basis of pseudoatomic orbitals with an orbital confining energy of 0.05 eV and an energy cutoff of 150 Ry were used, with Perdew-Burke-Ernzerhof functional [58]. The k-point sampling contains 6 k-points along the spiral axis (simulation cell has length $b$ along spiral axis). Spirals were optimized using Hellmann-Feynman forces down to 0.01 eV/Å tolerance[60]. Because of an elastic axial stress, the relaxation resulted in an average layer separation of $b \approx 3.2$ Å, somewhat smaller than graphite interlayer distance. Therefore, because van der Waals forces are much weaker than elastic forces, and because their role for the electronic structure is insignificant, they could be safely neglected. Also discussion in this work weakly might be affected by the fact that super cell optimization shows a possible for smallest spiral to change a hexagonal base to a pentagonal. For VASP calculations, the projector augmented wave and generalized gradient approximation for exchange and correlation energy were used.[58] Kohn-Sham orbitals were expanded in plane-wave basis set with energies up to 550 eV and the Brillouin zone was sampled over $1 \times 1 \times 4$ Monkhorst-Pack grid. For the pseudopotentials, Vanderbilt's ultrasoft potentials [61] with cutoff energy of 58 Ry was used. All used first-principle methods gave results in excellent agreement.

- **Acknowledgment** SMA is thankful Parijat Sengupta (Purdue, ECE) for fruitful discussion and Purdue, MSE for financial support. PK and CGR acknowledges the Academy of Finland for funding.

- **Correspondence** Correspondence and requests for materials should be addressed to SMA: (email: savdoshe@purdue.edu)

---

# Supplementary Materials for "Topological Signatures in the Electronic Structure of Graphene Spirals"

S.M. Avdoshenko,[1, *] P. Koskinen,[2] H. Sevinçli,[3] A.A. Popov,[4] and C.G. Rocha[2]

[1]*School of Materials Engineering, Purdue University, West Lafayette, Indiana, USA*

[2]*Nanoscience Center, Department of Physics,*
*University of Jyväskylä, 40014 Jyväskylä, Finland*

[3]*DTU Nanotech, Department of Micro- and Nanotechnology,*
*Technical University of Denmark, DK-2800, Denmark*

[4]*The Leibniz Institute for Solid State and Materials Research in Dresden, 01069 Dresden*

(Dated: November 13, 2012)

PACS numbers:

[*]Electronic address: savdoshe@purdue.edu

The figures 1, 2, 3 and the section *Hamiltonian transformation in response to curvature* are meant to support the explanations described in the main manuscript entitled "Topological Signatures in the Electronic Structure of Graphene Spirals".

## A. Hamiltonian transformation in response to curvature

In this section, we illustrate the relation between spatial curvature and effective fields. The curvature adds one extra dimension in the system in the same way as occurs in graphene when it turns into spirals. In general terms, Rashba spin-orbit coupling is similar to the Coriolis term describing spinning particles in a rotating reference frame [1–3]. Therefore, a Hamiltonian describing a rotating particle in an one-dimensional helix can acquire the same form as Rashba Hamiltonian. This could also be verified for the case of spinning polarized light which has its physics directly related to spin-orbit interaction [4].

Let us assume a toy-system such as the one depicted in Fig. 1 (a): two sites per unit cell enclosed in the rectangle, each of them being described by the on-site energy $\epsilon$. They are connected via the hopping parameter $\tau$ following a sort of crossing configuration as shown in the figure. Based on this topology, the 2×2-Hamiltonian $\mathbf{h_x}$ (**x**-means cross configuration) describing the electronic structure of this system can be written as

$$\mathbf{h_x} = \begin{pmatrix} \epsilon + \tau \cos(ka) & \tau + \tau \cos(ka) \\ \tau + \tau \cos(ka) & \epsilon + \tau \cos(ka) \end{pmatrix}$$

where $k$ is the wave vector and $a$ is the translational size of the unit cell. The system now undergoes a certain spatial deformation which brings the atoms located in the lower line closer as it is shown in Fig. 1(b). Under this transformation, the new unit cell assumes a trapezoid shape due to the shrinking of the lower atomic chain. This simple distortion mimics the case of graphene spirals since there its perimeter enclosing the inner edge is considerably shorter in comparison to the outer edge. Under this transformation, we assume that the strength in the chain is minimal and only off-diagonal elements can be affected by the distortion.

From Fig. 1(b), one can notice that the connectivity among the atoms in the upper chain is lost and a new 2×2-Hamiltonian, $\mathbf{h_s}$ (**s**-means spiral), emerges. The diagonal elements are preserved however the off-diagonal parts experience the curvature influence described by the element $\mathbf{f}$. The Hamiltonian can be then re-written as

$$\mathbf{h_s} = \begin{pmatrix} \epsilon + \tau cos(ka) & \tau + (\tau - \mathbf{f})cos(ka) \\ \tau + (\tau - \mathbf{f})cos(ka) & \epsilon + \tau cos(ka) \end{pmatrix}$$

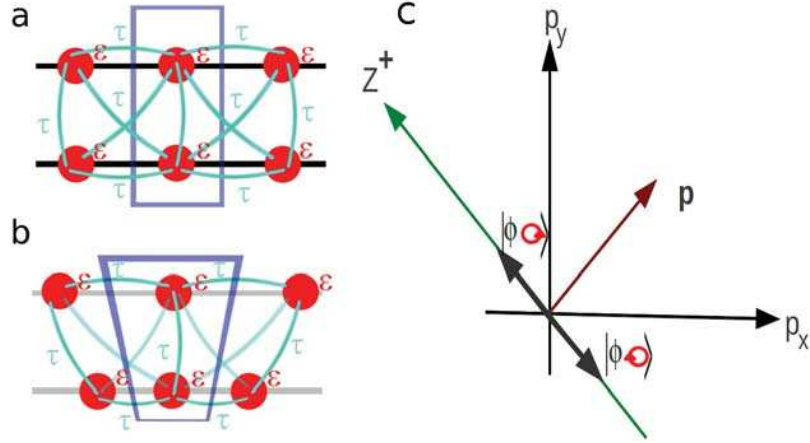

FIG. 1: Representation of an atomic chain with two sites per unit cell with (a) crossing coupling configuration and (b) after the distortion of the sites in the lower line. This transformation represents a conformal mapping from a rectangular into a trapezoid cell. (c) In the case of a spiral with its translational direction along z-axis, its divergence in the potential makes the contributions coming from z-direction more robust and the wave functions are now distinguished by clockwise and anti-clockwise orientations.

This modified Hamiltonian resembles the Rashba term when it is written in terms of $\mathbf{h_x}$, i.e. $\mathbf{h_s} = \mathbf{h_x} + (\mathbf{f}/\tau)\hat{\sigma}_x \times \hat{p}$, where $\mathbf{f}/\tau$ represents the curvature coupling, and $\hat{\sigma}_x$, $\hat{p}$ are the Pauli matrix and the momentum, respectively. Adapting this transformation to spirals where their tri-dimensional moment-space has orthogonal $p_x, p_y$-components [see Fig. 1(c)], we can write

$$\mathbf{h_s} = \begin{pmatrix} \epsilon + \tau \cos(ka) + \tau \sin(ka) & (\tau - \mathbf{f}) \sin(ka) + (\tau - \mathbf{f}) \cos(ka) \\ (\tau - \mathbf{f}) \sin(ka) + (\tau - \mathbf{f}) \cos(ka) & \epsilon + \tau \cos(ka) + \tau \sin(ka) \end{pmatrix}$$

Further manipulation of $\mathbf{h_s}$ will lead to the two-dimensional Rashba Hamiltonian form $\mathbf{h_s} = \mathbf{h_u} + (\mathbf{f}/\tau)\hat{\sigma}_x \times \hat{p}_y - (\mathbf{f}/\tau)\hat{\sigma}_y \times \hat{p}_x$ being $\mathbf{h_u}$ the Hamiltonian of the unfolded spiral. Likewise

Rashba term, this Hamiltonian affects the Fermi surface of the system but it is not capable of producing a spin polarized pure state which must be included explicitly.

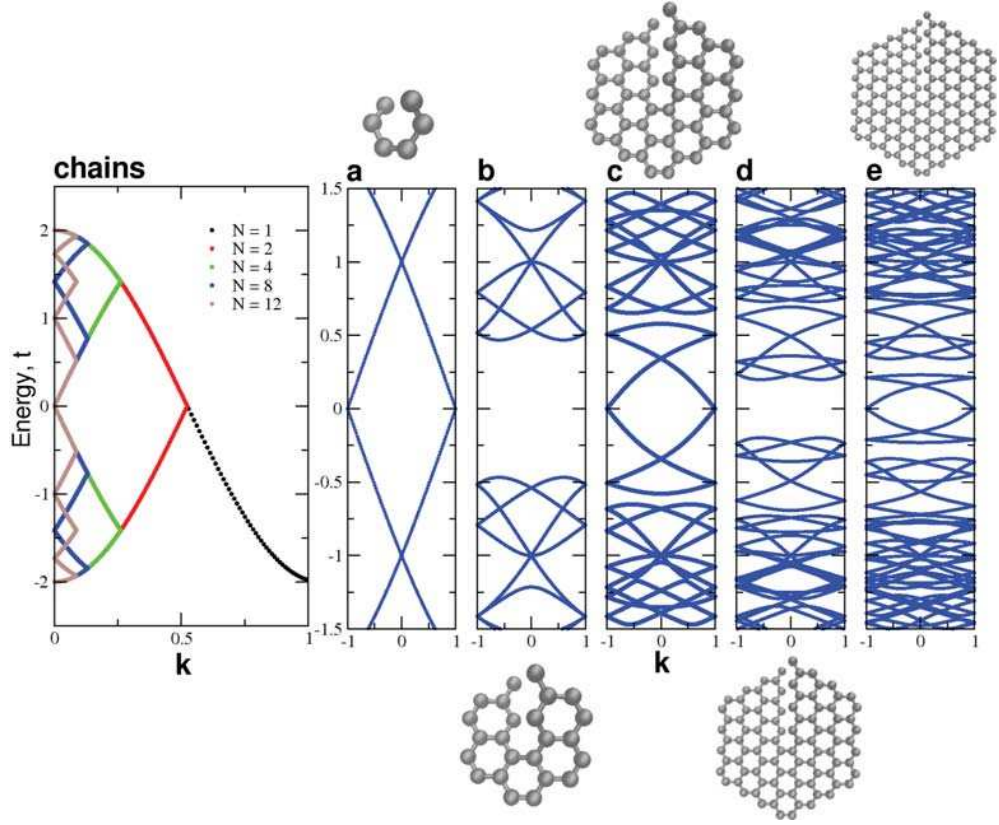

FIG. 2: (Chains) Band structures obtained for a linear atomic chain via simple tight-binding picture. The number of atoms in the unit cell (with lattice length of $a$) varies from $N = 1$ to $N = 12$. Energies are plotted in units of hopping parameter $t$ and the Fermi level is set at zero. Band structures of graphene spirals obtained via simple tight-binding model. (a) ac[1]zz[0], (b) ac[1]zz[0-1], (c) ac[1]zz[0-2], and (d) ac[1]zz[0-3]. Energies are shown in units of hopping parameter $t$ and Fermi energy is set at zero. Vertical axes span the $k$-vector intervals $[-\pi/b, \pi/b]$, where $b$ is the interlayer separation defining the periodicity of one full turn in the spiral.

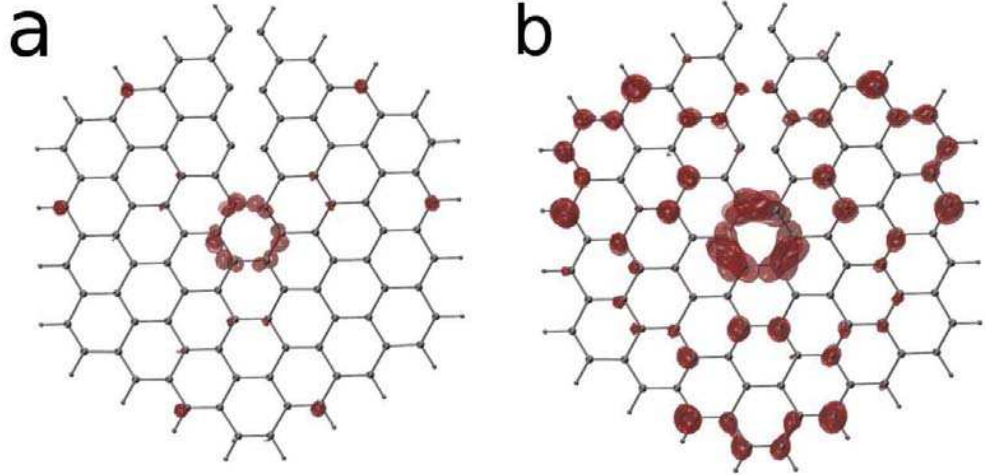

FIG. 3: (a-b) Different (0.1, 0.01) isosurface for local density of states at the Fermi level for the graphene spiral ac[1]zz[0-2] obtained from density functional theory implemented within SIESTA package[5].